\documentclass[journal]{IEEEtran}
\usepackage{algorithmic}
\usepackage{algorithm}

\usepackage{cite}       

\usepackage{graphicx}   

\usepackage{subfigure}  

\usepackage{url}        

\usepackage{amssymb}   
\usepackage{amsmath}   

\def\env{\epsilon}
\def\fenv{\hat{\epsilon}}
\def\noise{\xi}

\def\mismatx{\hat{x}_{\mathrm{mis}}}
\def\matx{\hat{x}_{\mathrm{mat}}}
\def\SNR{\mathrm{SNR}}
\def\hmat{h_m}

\def\Bmat{{B_\mathrm{mat}}}
\def\Bmis{{B_\mathrm{mis}}}
\def\xmis{{\hat{x}_\mathrm{mis}}}
\def\xmat{{\hat{x}_\mathrm{mat}}}

\def\N{\mathbb{N}}
\def\Z{\mathbb{Z}}

\newcommand{\E}[1]
{
  \mathrm{E}\,#1
}

\newcommand{\Var}[1]
{
  \mathrm{Var}\,#1
}

\begin{document}

\title{General radar transmission codes that minimize measurement error of a static target}

\author{Juha~Vierinen,~Markku~Lehtinen,~Mikko~Orisp\"a\"a,~and~Baylie~Damtie
\thanks{Manuscript received January 20, 2007; revised November 18, 2007.
        The authors are affiliated with Sodankyl\"a Geophysical
        Observatory, correspondence: \{j@sgo.fi, markku.lehtinen@sgo.fi\}}
}

\markboth{IEEE TRANSACTIONS ON INFORMATION THEORY,~Vol.~1, No.~11,~November~2007}{Vierinen \MakeLowercase{\textit{et al.}}: Arbitrarily modulated radar codes that minimize measurement error of a stationary target}

\maketitle

\begin{abstract}
The variances of matched and sidelobe free mismatched filter
estimators are given for arbitrary coherent targets in the case of
aperiodic transmission. It is shown that mismatched filtering is often
better than matched filtering in terms of estimation accuracy. A
search strategy for finding general transmission codes that minimize
estimation error and satisfy constraints on code power and amplitude
range is then introduced. Results show that nearly perfect codes, with
performance close to a single pulse with the same total power can be
found. Also, finding these codes is not computationally expensive and
such codes can be found for all practical code lengths. The estimation
accuracy of the newly found codes are compared to binary phase codes
of similar length and found to be better in terms of estimator
variance. Similar transmission codes might be worth investigating also
for sonar and telecommunications applications.
\end{abstract}


\IEEEpeerreviewmaketitle

\section{Introduction}
\PARstart{P}{hase} modulation of a radar transmission is a well known
method for increasing radar transmission power, while still
maintaining a good range resolution. Such transmission codes can
consist of two or more individual phases. The performance of binary,
quadri and polyphase codes has been thoroughly
inspected in terms of heuristic criteria, such as the integrated sidelobe level (ISL), or peak to sidelobe level (PSL) \cite{Barker,Nunn1,Turyn,Turyn2,Lindner,Taylor,Mow}.  In
previous work, binary phase codes have also been evaluated in terms of
estimation accuracy of a static target, when using an optimal
sidelobe free mismatched filter for periodic \cite{Key,Rohling,Luke} and 
aperiodic signals \cite{Lehtinen}.

We first examine the behaviour of matched and optimal sidelobe free
mismatched filter estimators for a point like and a uniform target. In
the case of a point-like target, we get the well known result that the
matched filter is optimal, and the sidelobe free mismatched filter
has a larger estimator variance, which depends mainly on the sidelobe
power, and is thus not necessarily very high. In the case of a uniform
target, we see that the matched filter produces biased results and in
addition to the bias, it also has a worse estimator variance in many cases. 
(Here we consider the mean value of the error term as bias and call the second moments
of the error term around the mean the estimator variance).

\section{General transmission code}

A code with length $L$ can be described as an infinite length sequence
with a finite number of nonzero pulses with phases and amplitudes
defined by parameters $\phi_k$ and $a_k$. These parameters obtain
values $\phi_k \in [0,2\pi]$ and $a_k \in
[a_{\mathrm{min}},a_{\mathrm{max}}]$, where $k \in [1,\ldots, L] : k \in
\N$. The reason why one might want to restrict the amplitudes to some
range stems from practical constraints in transmission
equipment. In most traditional work, the amplitudes have been set to $1$ and
often the number of phases has also been restricted, eg., in the case
of binary phase codes to $\phi_k \in \{0,\pi\}$.

Defining $\delta(t)$ with $t \in \Z $ as
\begin{equation}
\delta(t) = \left\{
\begin{array}{lcr}
1  & \mbox{when}      & t = 0 \\
0  & \mbox{otherwise} &
\end{array} \right.
\end{equation}
we can describe an arbitrary baseband radar code $\env(t)$ as
\begin{equation}
\env(t) = \sum_{k=1}^{L} a_k e^{i\phi_k} \delta(t - k + 1).
\end{equation}

In addition to this, we restrict the total transmission code power to be
constant for all codes of similar length. Without any loss of
generality, we set code power equal to code length
\begin{equation}
L = \sum_{t=1}^{L} |\env(t)|^2.
\end{equation}
This will make it possible to compare estimator variances of codes
with different lengths and therefore different total transmission
powers. Also, it is possible to compare codes of the same length and
different transmission power simply by treating $L$ as transmission
power.

\section{Measurement equation}

Equation \ref{measeq} describes the basic principle of estimating a
 coherent radar target\footnote{scattering amplitude stays while the
 transmission passes the target} using a linear filter. When the
 target is assumed to be infinite length and using roundtrip time as
 range, the scattering from a target is simplified to convolution of
 the transmission with the target. In this convolution equation,
 $m(t)$ denotes the measured signal, $\sigma(t)$ denotes the unknown
 target, $\env(t)$ denotes the transmitted waveform and $\noise(t)$
 represents thermal noise, which is assumed to be Gaussian white noise
 with power $\SNR^{-1}$. Finally, $h(t)$ represents the decoding
 filter used to decode the signal, it can be eg., a matched or
 mismatched filter.
\begin{equation}
m(t) = \left[ \sigma(t) \ast \env(t) + \noise(t) \right]\ast h(t)
\label{measeq}
\end{equation}

Assuming that the Fourier transformation of the transmitted waveform
contains no zeros, a solution to the previous equation can be found
easily in frequency domain \cite{Lehtinen}. Using notation
$\mathcal{F}_D\left\{\env(t)\right\} = \fenv(\omega)$ for a zero
padded discrete Fourier transform with transform length $M \gg L$, the
optimal sidelobe free mismatched filter can be defined as
$\mathcal{F}_D^{-1}\left\{L/\fenv(\omega)\right\} = \lambda(t)$. Such
a filter will be infinite length, but it is a mathematical fact that
the coefficients will exponentially approach zero 
\cite{Lehtinen2}, so one can use a truncated $\lambda(t)$ with
errors of machine precision magnitude. Also, it is known that
filtering with $\lambda(t)$ is the minimum mean square estimator for
target amplitude. 

In the case of the mismatched filter, we set $h(t) = \lambda(t)$ in
the measurement equation, which can be simplified into the following
form

\begin{equation}
m_{\lambda}(t) = L\sigma(t) + \underbrace{(\noise \ast \lambda)(t)}_{\mathrm{measurement\ error}}.
\label{filter}
\end{equation}

In the case of a matched filter $\hmat(t) = \overline{\env(-t)}$, one
can also extract the target from the measurement equation. From
equation $(\env * \hmat)(t) - r(t) = L\delta(t)$, we see that the
matched filter can be expressed using the mismatched filter
$\lambda(t)$ and code autocorrelation function sidelobes $r(t)$ as
\begin{equation}
\hmat(t) = \lambda(t) + \frac{1}{L}(\lambda * r)(t),
\label{matmismat}
\end{equation}
and thus we can write the matched filter measurement equation $m_m(t)$
as 
\begin{equation}
m_m(t) = L\sigma(t) + (r*\sigma)(t) + \underbrace{(\noise \ast \hmat)(t)}_{\mathrm{measurement\ error}}.
\label{filterm}
\end{equation}

\begin{figure}
\centering
\includegraphics[width=3in]{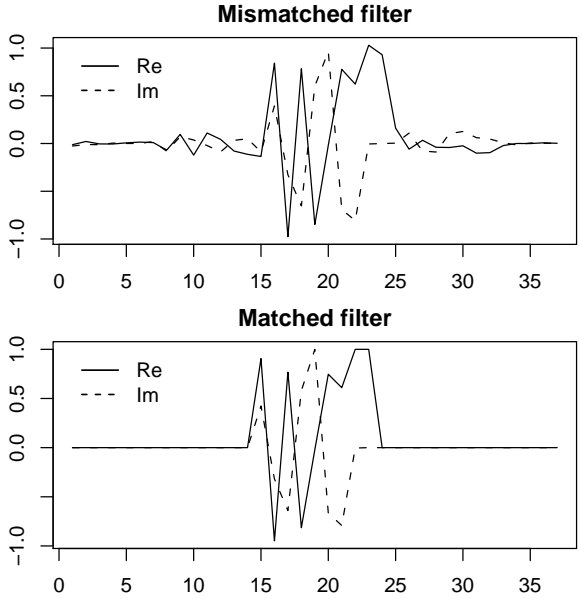}
\caption{The mismatched and matched filters of a polyphase code
$\phi_k \in [0,2\pi]$ with length $L=9$ and $R=0.974$.}
\label{filters}
\end{figure}

Equation \ref{matmismat} shows that the matched filter for a code with
integrated sidelobe power approaching zero
$\sum_{t=-\infty}^{\infty}|r(t)|^2 \rightarrow 0$ approaches the
sidelobe free mismatched filter $\hmat(t) \rightarrow \lambda(t)$.  In
this case measurement equations \ref{filter} and \ref{filterm} are the
same, which is a natural result. Figure \ref{filters} shows a
mismatched and a matched filter for a relatively good code.

\section{Estimators}


When estimating the power of a target, it is customary to use several
repetitions of a measurement. In this case, the target and the  thermal
noise are denoted as random variables, which are indexed with $q \in
\N$, ie., each repetition is a different random variable. The
measurement equation for repeated measurements is then written as 
\begin{equation}
m^q(t) = \left[ (\env*\sigma^q)(t) + \noise^q(t) \right]*h(t).
\label{measrep}
\end{equation}

Even though the scattering amplitude and thermal noise amplitude
change between measurements, we assume that the statistical properties
of the thermal noise and the target are unchanged between
measurements, and this is what is estimated. The target is measured as
target power using sample variance, from which we subtract known bias
caused by the thermal noise entering the filter. The matched filter
target power estimator is thus
\begin{equation}
\xmat(t)  = - \frac{\Bmat}{L^2} + \frac{1}{N\,L^2} \sum_{q=1}^N |m_m^q(t)|^2
\end{equation}
and the mismatched filter 
\begin{equation}
\xmis(t)  = - \frac{\Bmis}{L^2} + \frac{1}{N\,L^2} \sum_{q=1}^N |m_{\lambda}^q(t)|^2.
\end{equation}

In these equations the thermal noise entering the filter is denoted
with $\Bmat = \SNR^{-1} \sum_{\tau=-\infty}^{\infty} |h_m(\tau)|^2$ and
$\Bmis = \SNR^{-1} \sum_{\tau=-\infty}^{\infty} |\lambda(\tau)|^2$. 

\section{Point-like target}

In baseband, the scattering from a point target is defined as a zero
mean complex Gaussian random process with the second moment defined
with the following expectation
\begin{equation}
\E{\sigma^q(t)\overline{\sigma^{p}(t')}} = x\delta(t-t_c)\delta(t'-t_c)\delta(q-p).
\end{equation}
In other words, the scattering is zero for all other ranges than
$t_c$, where the scattering power is $x$. Different repetitions are
not correlated.

In this case, it can be shown that the matched filter and mismatched
filter estimators are both unbiased, ie., $\E{\xmis(t)} = \E{\xmat(t)}
= x$. The estimator variances are:
\begin{equation}
\Var{\matx} = \frac{1}{N} \left(
x^2 + \frac{2\Bmat x}{L^2} + \frac{\Bmat^2}{L^4} \right)
\end{equation}
and 
\begin{equation}
\Var{\mismatx} = \frac{1}{N} \left(
x^2 + \frac{2\Bmis x}{L^2} + \frac{\Bmis^2}{L^4} \right).
\end{equation}

The target itself is a source of estimation errors, as it is a
Gaussian random variable (self-noise). The only code dependent terms are the
thermal noise terms $\Bmat$ and $\Bmis$. Thus, the only way to reduce
estimator variance is to reduce thermal noise. In the case of a
matched filter, the noise entering the filter is independent of the
code and proportional decoding filter power $L$. For a mismatched
filter, the thermal noise term is always larger than the matched
filter equalent, and it is highly code dependent. In order to compare
estimator performance, we can use the following ratio:
\begin{equation}
R =  \frac{\Bmat}{\Bmis} =
\frac{L}{\sum_{\tau=-\infty}^{\infty}|\lambda(\tau)|^2},
\label{bigr}
\end{equation}
which will approach $1$ when the performance of the optimal mismatched filter
approaches that of the matched filter. 

\section{Distributed target}

\begin{figure}
\centering
\includegraphics[width=3in]{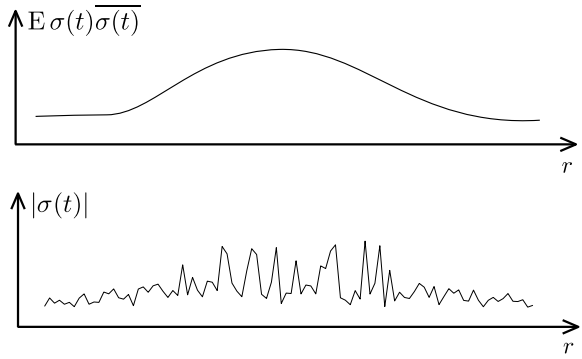}
\caption{The expectation of power and an example of an instance of the target.}
\label{distr}
\end{figure}

When the target is not point-like, the situation is different. A zero
mean time-stationary Gaussian scattering medium with power depending on range can be
defined as
\begin{equation}
\E{\sigma^q(t) \overline{\sigma^{p}(t')}} = x(t)\,\delta(t-t')\,\delta(q-p).
\end{equation}
Figure \ref{distr}. shows an example of $x(t)$ and the instantanious
scattering $\sigma(t)$.

In the case of a distributed target, it can be shown that the
expectation of the matched filter estimator is biased, with the
sidelobes convolved with the target. By defining the sidelobe term as
\begin{equation}
S(t) = \sum_{\tau=-\infty}^{\infty} |r(\tau)|^2 x(\tau - t),
\end{equation}
we can describe the matched filter estimator mean as
\begin{equation}
\E{\matx(t)} = x(t) + \frac{S(t)}{L^2}.
\end{equation}

On the other hand, the sidelobe free mismatched filter estimator is
unbiased. It has mean
\begin{equation}
\E{\mismatx(t)} = x(t).
\end{equation}

The variance of the estimators can also be found. The matched
filter has a variance 
\begin{align}
\Var{\matx(t)} = & \frac{1}{N} \Big[ x(t)^2 + \frac{2\Bmat\,x(t)}{L^2} +
  \frac{2S(t)x(t)}{L^2}  +  \nonumber \\ 
  & \frac{\Bmat^2}{L^4} + \frac{S(t)^2}{L^4} +
  \frac{2\Bmat\,S(t)}{L^4} \Big]
\label{matestvar}
\end{align}
 and the mismatched filter has variance:
\begin{equation}
\Var{\mismatx(t)} = \frac{1}{N}\left[
  x(t)^2 + \frac{2\Bmis\,x(t)}{L^2} + \frac{\Bmis^2}{L^4} \right]
\end{equation}

By inspecting these equations, one can see that the mismatched filter
variance is the same as it was for a point-like target, but the
matched filter has additional sidelobe terms. In many cases these
terms will cause the variance of the matched filter estimator to be
wider than the mismatched filter estimator variance. 

\begin{figure}
\centering
\includegraphics[width=3in]{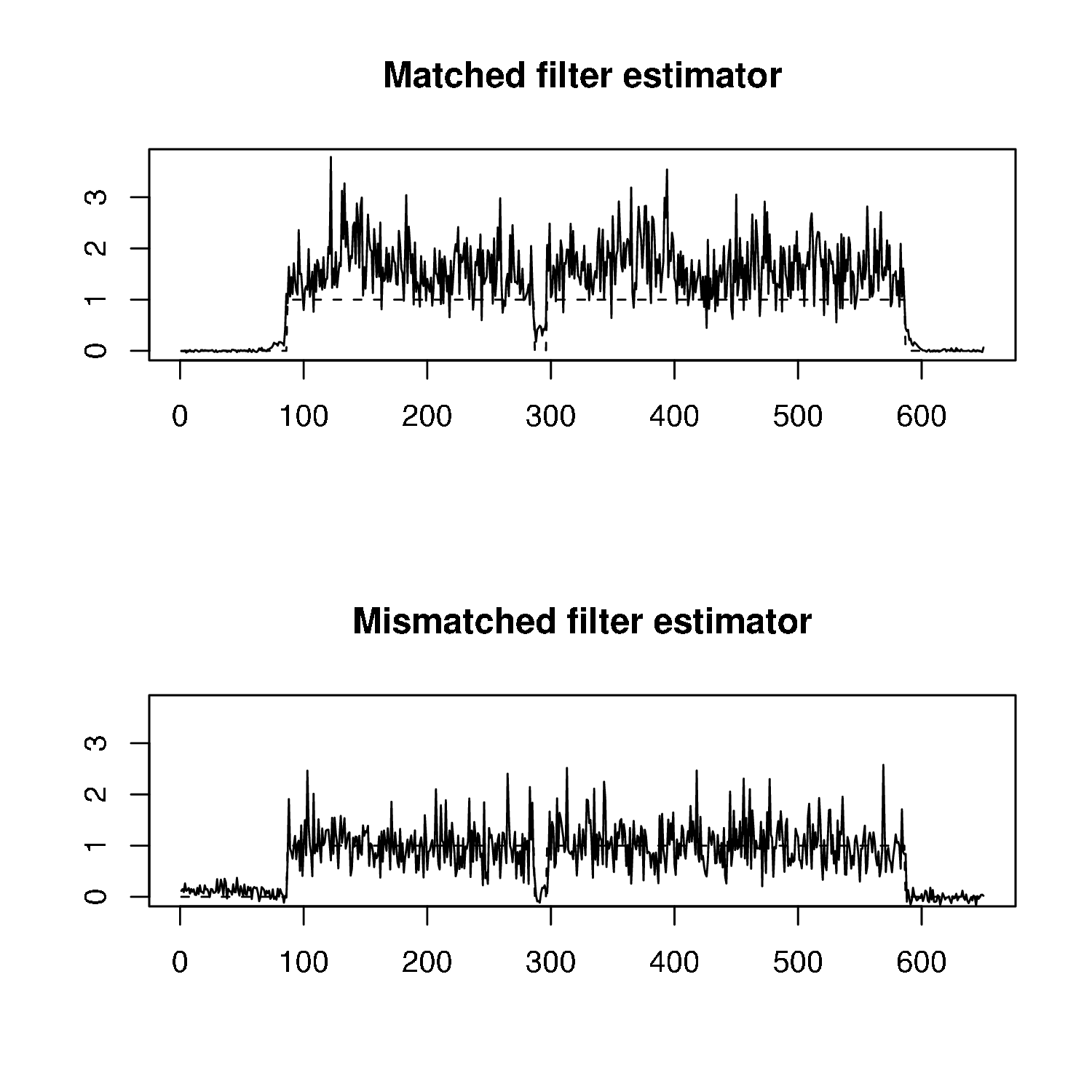}
\caption{Simulated uniform target estimated with a matched and
  mismatched filter using a random phase code. In this case the signal
  to noise ratio is one, and the code is relatively poor $R=0.23$. The
  matched filter estimator has bias and larger variance due to
  self-noise caused by the target. The correct target is zero
  everywhere else and $x(t)=1$ when $t \in [90,290] \cup [300,590]$.}
\label{simfig}
\end{figure}

Figure \ref{simfig} shows a simulated target that is probed with a
random phase code and then the target power is estimated with
matched and mismatched filter estimators. A relatively poor random
phase code with $R=0.23$ was used to emphasize the following
relevant features: 
\begin{enumerate}
\item With all but the smallest signal to noise ratios the matched
filter estimator has larger variance. For example, if the target is
assumed to be completely uniform $x(t) = 1$, the matched filter
estimator variance for the 13-bit Barker code is better only when
$\SNR < 0.05$. When the signal to noise ratio is higher than this, the
mismatched filter has better estimation variance. When $\SNR=1$, the
estimation variance of the mismatched filter is already $11$\% better
for the 13-bit Barker code. 

\item The matched filter has bias which depends on the sidelobes. For
  example, when the target is again uniform $x(t)=1$, the bias of the
  best binary phase codes of lengths $3$ to $42$ is around $0.1$, in
  other words, the target power estimate is $10$\% higher than it is
  in reality. In figure \ref{simfig}, the bias is about $80$\%.

\item The mismatched filter produces larger thermal noise. This can be
  seen on the outermost extremes in Figure \ref{simfig} where $x(t) =
  0$. This is code dependent, and depends on the value of $R$. When $R
  \rightarrow 1$, the thermal noise of a mismatched filter is equal to
  that of a matched filter.

\end{enumerate}

\begin{figure}
\centering
\includegraphics[width=3in]{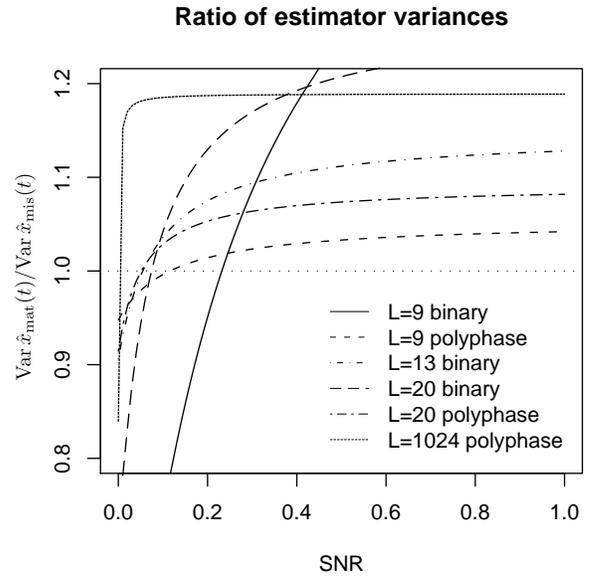}
\caption{The ratio of matched and mismatched filter estimator
  variances for a uniform target $x(t)=1$. The best performing
  polyphase and binary phase codes for several code lengths is
  shown. When the ratio is $>1$, the mismatched filter estimator is
  better.}
\label{estratios}
\end{figure}

Figure \ref{estratios} shows the ratio of matched and mismatched
filter variances for the best polyphase and binary phase codes of
several different lengths as a function of signal to noise
ratio. When the ratio is smaller than one, the matched filter performs
better. It can be seen from the figure that when $\SNR$ is increased,
the mismatched filter is better after some threshold $\SNR$, and the
ratio of variances asymptotically approaches a certain code dependent
ratio. Also, when code length is increased, the threshold $\SNR$ where
the mismatched filter has better variance is lowered. This can be seen
from the behaviour of the polyphase code of length 1024.

\section{Code optimality}

In our considerations, we only concentrate on minimizing the
mismatched filter estimator variance, because the matched filter is
biased by the code sidelobes and also often has larger estimator variance for a distributed
target. In any case, it is possible to inspect matched filter
estimator performance by using equation \ref{matestvar}.

From the equations of mismatched filter estimator variance it is clear
that the code affects estimation variance. The estimator variance is
the same for both distributed and point targets, so it is sufficient
to maximize the ratio $R$ described in equation \ref{bigr}. But what
does maximizing $R$ mean? From eq. \ref{matmismat}, which describes a
matched filter in terms of a mismatched filter and matched filter ACF
sidelobes $r(t)$, one can see that when sidelobe power
$p=\sum_{t=-\infty}^{\infty}|r(t)|^2$ approaches zero the mismatched
filter approaches the matched filter

\begin{equation}
\lim_{p \rightarrow 0} \hmat(t) = \lambda(t).
\end{equation}

In this case we have a code with $R=1$, ie., the matched and
mismatched filters are the same and ACF is a single spike $\env(t) * \overline{\env(-t)} =
L\delta(t)$. Therefore, even though we are restricting ourselves to
the mismatched filter, the same codes will also be good when used as a
matched filter. The closer $R$ is to $1$, the smaller the sidelobes
and thus matched filter error.

Traditional code optimality criteria also reflect code goodness, but
their relation to mismatched filter estimation accuracy is not that
well defined. Still, it is evident from equation \ref{matmismat} that
the sidelobes of the code autocorrelation function directly affect the
performance of the mismatched filter by making the filter longer than
the matched filter, allowing more thermal noise to enter the
estimate. Thus, traditional code optimality criteria such as peak to
maximum sidelobe level (PSL) or code power divided by integrated
sidelobe power (MF) will also reflect code goodness. In the limiting
case, when $PSL \rightarrow 0$ and $MF \rightarrow \infty$ it is clear
that $R$ will also have limit $R \rightarrow 1$.

\section{Code search algorithm}

\begin{algorithm}[h!]
\begin{algorithmic}
\REPEAT
\FOR{$b = 1$ to $\mathrm{code\ length}$}
\STATE $\phi[b] \Leftarrow \mathrm{Uniform Random}(0,2\pi)$
\STATE $a[b] \Leftarrow 1$
\STATE $\env[b] \Leftarrow a[b]*\exp(j\phi[b])$
\ENDFOR
\FOR{$i = 1$ to $\mathrm{number\ of\ iterations}$}

\IF{$\mathrm{Uniform Random}(0,1) < 0.5$} 

\STATE $b \Leftarrow \mathrm{floor}(\mathrm{Uniform
  Random}(0,\mathrm{code\ length})) + 1$
\STATE $\Delta\phi \Leftarrow \mathrm{Uniform Random}(0,2\pi)$
\STATE $\mathrm{old}\env \Leftarrow \env[b]$
\STATE $\env[b] \Leftarrow \exp(j\Delta\phi)$
\STATE $NewR \Leftarrow \mathrm{CalculateR}(c)$
\IF{$NewR > R$} 
\STATE $R \Leftarrow NewR$
\STATE $\phi[b] \Leftarrow \Delta\phi$
\ELSE
\STATE $\env[b] \Leftarrow \mathrm{old}\env$
\ENDIF

\ELSE

\STATE $b1 \Leftarrow \mathrm{floor}(\mathrm{Uniform Random}(0,\mathrm{code\ length})) + 1$
\STATE $b2 \Leftarrow \mathrm{floor}(\mathrm{Uniform Random}(0,\mathrm{code\ length})) + 1$
\STATE $\Delta{}a \Leftarrow \mathrm{Normal Random}(0,1)$
\STATE $\mathrm{olda} \Leftarrow a$
\STATE $\mathrm{old}\env \Leftarrow \env$
\STATE $q \Leftarrow 4a[b2]^2-8a[b1]\Delta{}a - 4\Delta{}a^2$
\IF{$q > 0\,\,\mathbf{and}\,\,b1 <> b2$}
\STATE $\Delta{}a2 \Leftarrow -a[b2]-0.5\sqrt{q}$
\STATE $a[b1] \Leftarrow a[b1] - \Delta{}a$
\STATE $a[b2] \Leftarrow a[b2] + \Delta{}a2$
\STATE $c[b1] \Leftarrow a[b1]\exp(j\phi[b1])$
\STATE $c[b2] \Leftarrow a[b2]\exp(j\phi[b2])$
\STATE $NewR \Leftarrow \mathrm{CalculateR}(\env)$
\IF{$NewR > R$} 
\STATE $R \Leftarrow NewR$
\ELSE
\STATE $\env \Leftarrow \mathrm{old}\env$
\STATE $a \Leftarrow \mathrm{olda}$
\ENDIF
\ENDIF

\ENDIF

\ENDFOR
\UNTIL{R good enough}
\end{algorithmic}
\caption{Random local improvement algorithm.}
\label{rwopt}
\end{algorithm}

\begin{figure}
\centering
\includegraphics[width=3in]{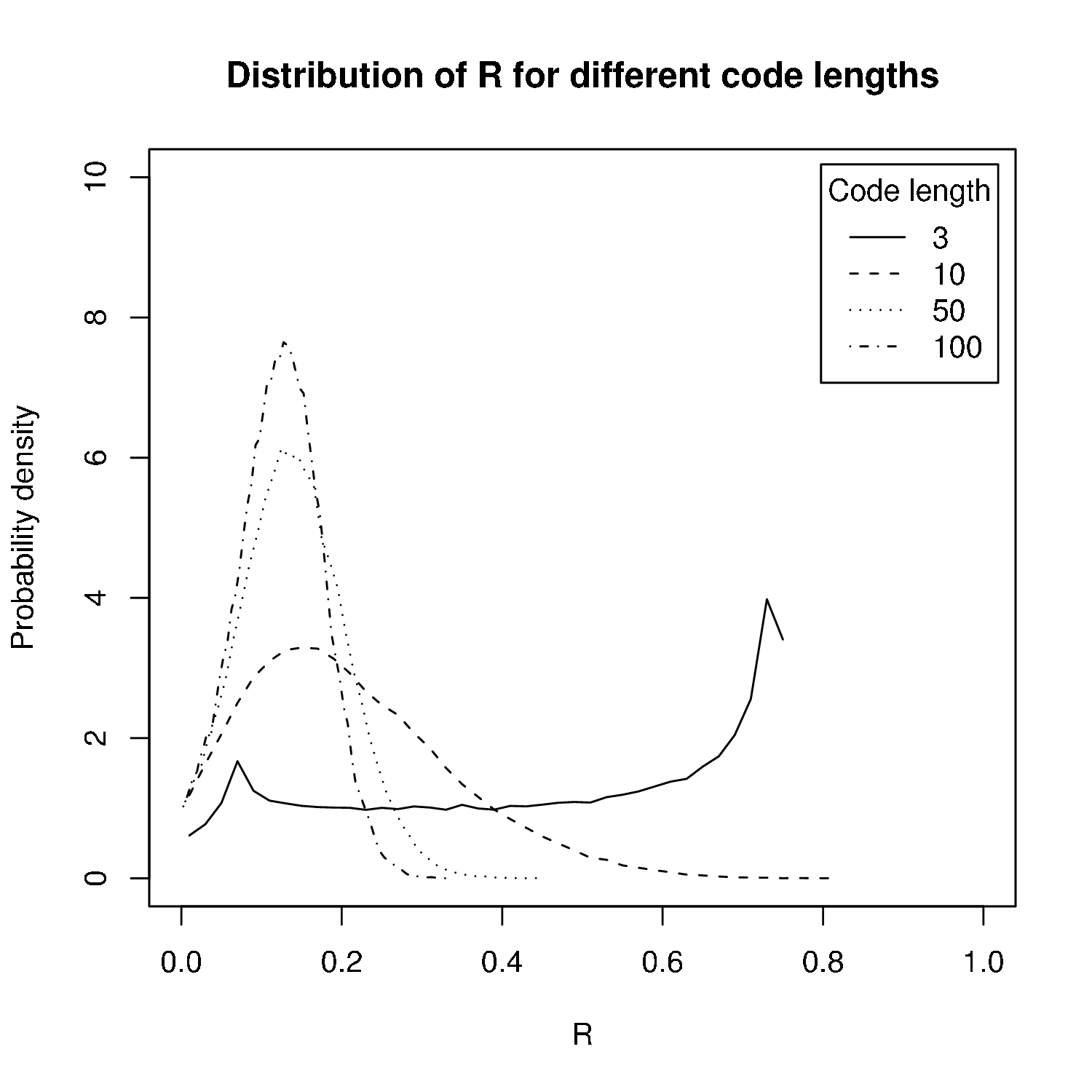}
\caption{Distribution of R for random polyphase codes, ie., $a_k=1$
  and $\phi_k \in [0,2\pi]$ for all $k$.}
\label{codedist}
\end{figure}

Lacking an analytic method of obtaining codes with $R$ close to one,
while statisfying the constraint on code amplitude range $a_k \in
[{a_{\mathrm{min}},a_{\mathrm{max}}}]$, we resort to numerical
means. In order to get an overview of how the performance of codes is
distributed among codes, we sampled several code lengths using $10^6$
randomly chosen polyphase codes (constant amplitude), and used a
histogram to come up with an estimate distribution of R. This shown in
figure \ref{codedist}. It is evident that as the code length grows, it
becomes nearly impossible to find good codes by searching them in a
purely random fashion. Therefore, in order to proceed numerically,
some form of optimization algorithm was needed.

We used a heuristic optimization algorithm specifically created for
this task, with the purpose of robustly converging to a maxima of $R$
as a function of a code, while satisfying constraints on code
amplitude range. The code is described in algorithm \ref{rwopt}. The
idea is as follows:

\begin{enumerate}
\item We first generate a code with all bauds at random phases and
  unit amplitudes. 
\item For a fixed amount of iterations, a new phase or amplitude is
randomized for a randomly selected baud, and $R$ calculated for the
resulting trial code. If the amplitude is changed, we also select
another baud and change its amplitude in the opposite direction in
order to maintain total code power at $L$. If the code is good enough,
we select it as our new current working code. 
\item After each ``optimization run'', we will find a code at some
local maximum. The optimization runs (Step 2.) are then repeated with
new random initial code until a satisfactory result has been obtained.
\end{enumerate}

The number of iterations of an optimization run is a tunable parameter
of the algorithm, it varies from $10^3$ for small code lengths to
$10^6$ for codes with length $L > 10^3$. 

One of the main reasons for robustness of this algorithm is that it
does not follow the largest gradient, but instead follows a random
positive gradient, making it more likely that more local maximas of
$R$ are visited.

The algorithm has also been applied with some modifications for more
resticted cases, such as binary and quadriphase codes that are too
long to search exhaustively.

\section{Search results}

We applied the search algorithm for code lengths $3$ to $4096$ using
three different amplitude ranges: $A_1 := [1,1]$, $A_2 := [0.95,1.05]$
and $A_3 := [0,2]$. The first of these is a polyphase code with
constant amplitude, the other two allow a certain amount of
amplitude deviation around 1. Results are shown in table \ref{res} as the best
value of $R$ found for given code length and amplitude range. For
comparison, the values of best binary phase codes are also shown in
column $B$. Some selected codes are given in table
\ref{rescodes}.\footnote{The software and more complete results can be
found at \url{http://mep.fi/mediawiki/PhaseCodes}}.

\begin{table}
\renewcommand{\arraystretch}{1.0}
\caption{Best transmission codes found.}
\label{res}
\centering
\begin{tabular}{c|c|c|c|c}
\hline
Length & $A_1$ & $A_2$ & $A_3$ & $B$ \\
\hline
3 & 0.745 & 0.775 & 1.000 & 0.745 \\ 
4 & 0.679 & 0.748 & 1.000 & 0.679 \\ 
5 & 0.866 & 0.900 & 1.000 & 0.866 \\ 
6 & 0.676 & 0.743 & 1.000 & 0.676 \\ 
7 & 0.894 & 0.917 & 1.000 & 0.705 \\ 
8 & 0.817 & 0.862 & 1.000 & 0.756 \\ 
9 & 0.974 & 0.979 & 1.000 & 0.618 \\ 
10 & 0.886 & 0.921 & 1.000 & 0.678 \\ 
11 & 0.926 & 0.946 & 1.000 & 0.804 \\ 
12 & 0.899 & 0.927 & 1.000 & 0.853 \\ 
13 & 0.954 & 0.971 & 1.000 & 0.952 \\ 
14 & 0.926 & 0.948 & 1.000 & 0.835 \\ 
15 & 0.951 & 0.968 & 1.000 & 0.870 \\ 
16 & 0.937 & 0.958 & 1.000 & 0.788 \\ 
17 & 0.953 & 0.969 & 1.000 & 0.773 \\ 
18 & 0.927 & 0.954 & 1.000 & 0.792 \\ 
19 & 0.968 & 0.958 & 1.000 & 0.831 \\ 
20 & 0.956 & 0.973 & 1.000 & 0.838 \\ 
21 & 0.962 & 0.976 & 1.000 & 0.835 \\ 
22 & 0.956 & 0.974 & 1.000 & 0.806 \\ 
23 & 0.968 & 0.983 & 1.000 & 0.824 \\ 
24 & 0.959 & 0.974 & 1.000 & 0.835 \\ 
25 & 0.968 & 0.982 & 1.000 & 0.853 \\ 
26 & 0.960 & 0.976 & 1.000 & 0.877 \\ 
27 & 0.953 & 0.973 & 1.000 & 0.862 \\ 
28 & 0.956 & 0.970 & 1.000 & 0.847 \\ 
29 & 0.959 & 0.974 & 1.000 & 0.853 \\ 
30 & 0.940 & 0.971 & 1.000 & 0.864 \\ 
31 & 0.950 & 0.976 & 1.000 & 0.860 \\ 
32 & 0.971 & 0.971 & 1.000 & 0.843 \\ 
33 & 0.982 & 0.973 & 1.000 & 0.856 \\ 
34 & 0.940 & 0.976 & 1.000 & 0.867 \\ 
35 & 0.961 & 0.979 & 1.000 & 0.851 \\ 
36 & 0.948 & 0.976 & 1.000 & 0.847 \\ 
37 & 0.941 & 0.978 & 1.000 & 0.850 \\ 
38 & 0.948 & 0.969 & 1.000 & 0.855 \\ 
39 & 0.953 & 0.970 & 1.000 & 0.849 \\ 
40 & 0.959 & 0.981 & 1.000 & 0.842 \\ 
41 & 0.940 & 0.971 & 1.000 & - \\ 
42 & 0.960 & 0.970 & 1.000  & - \\
64   & 0.966 & -     & 0.999 & - \\
128  & 0.941 & -     & 0.999 & - \\
256  & 0.946 & -     & 0.998 & - \\
512  & -     & 0.944 & 0.998 & - \\
1028 & -     & 0.929 & 0.997 & - \\
2048 & -     & 0.930 & 0.996 & - \\
4096 & -     & 0.929 & 0.995 & - \\
\hline
\end{tabular}
\end{table}

The results show that polyphase codes are better than binary phase codes.
When we allow the amplitude of the code to change, we get still better
codes. Nearly all of the codes with the largest amplitude range $A_3$
have performance comparable to that achievable with complementary
codes. In this case $R$ is less than $0.5 \cdot 10^{-4}$ from
theoretical maximum.

\begin{figure*}
\centering
\includegraphics[width=\textwidth]{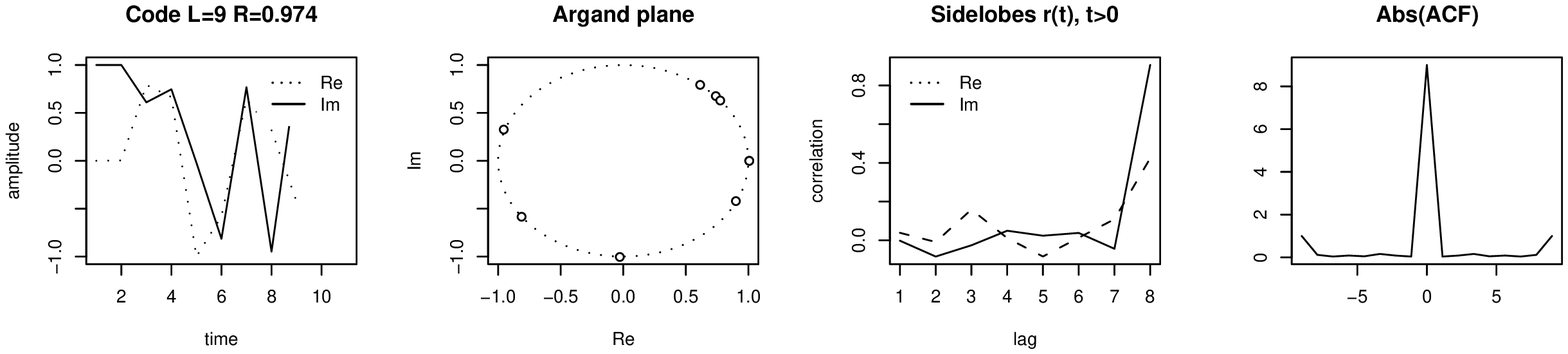}
\caption{The best polyphase code with $R=0.974, L=9$. This
  code also has exceptionally high merit factor of 25.6. It is evident
  from the ACF, that this is nearly the most optimal code possible for
  a code with unit norm amplitudes, as the outermost extremes of a
  single pulse have to be 1. The phases in degrees are: $0, 0.2, 52.4, 41.7, -91.1, -144.5, 39.8, 161.4, -25.1$.}
\label{acfL9}
\end{figure*}

\begin{figure*}
\centering
\includegraphics[width=\textwidth]{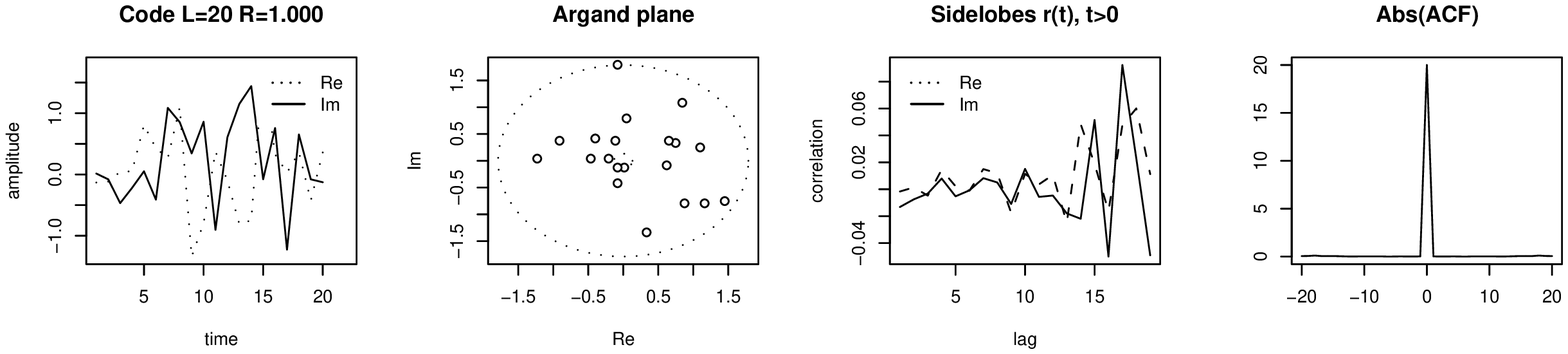}
\caption{An amplitude and phase modulated code with amplitudes in
  range $a_k \in [0,2]$. It can be seen that the sidelobes can be
  reduced to nearly zero, when amplitude modulation is allowed.}
\label{amfm20}
\end{figure*}

\begin{figure*}
\centering
\includegraphics[width=\textwidth]{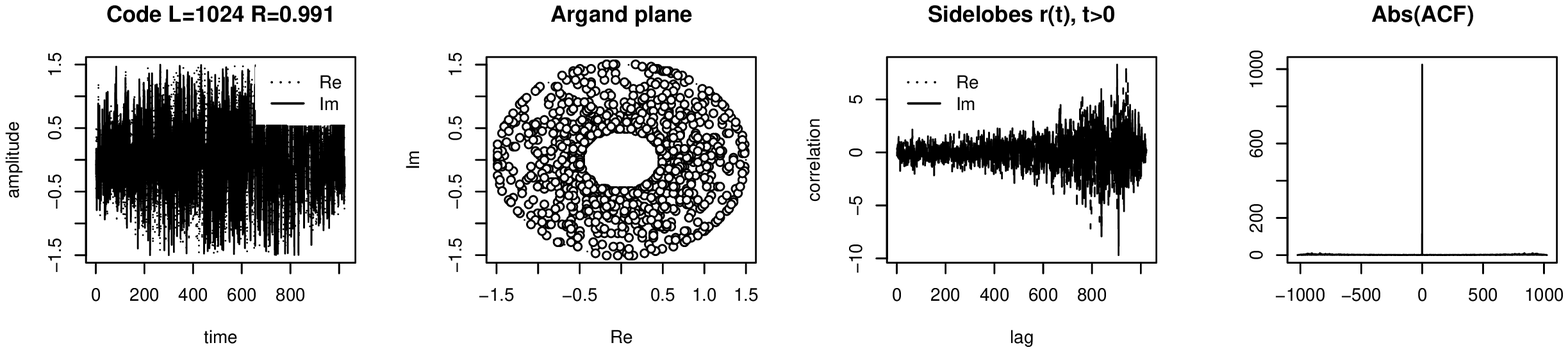}
\caption{A long code with amplitudes in range $a_k \in [0.5,1.5]$.}
\label{amfm1024}
\end{figure*}

Figures \ref{acfL9}, \ref{amfm20} and \ref{amfm1024} show several of
these codes. The first one is the best polyphase code of length
$9$. It is interesting because it has nearly optimal shape of ACF.
(The values of the ACF for lags $\pm 8$ are necessarily of norm one, because the 
first and the last element of the code have
norm one, but the rest of the ACF values are close to zero). 
The second figure shows an amplitude
and phase modulated code and the third shows a longer code of length
1024 with more restricted amplitudes.

\begin{table*}
\renewcommand{\arraystretch}{1.1}
\caption{Some selected codes}
\label{rescodes}
\centering
\begin{tabular*}{\textwidth}{ @{\extracolsep{\fill}} |c|c|p{0.4\textwidth}|p{0.4\textwidth}|}
\hline
Length & R & Phases $\phi_k$ (degrees) & Amplitudes $a_k$ \\
\hline

9  & 0.973 & 0 0.2 52.4 41.7 -91.1 -144.5 39.8 161.4 -25.1 & 1 1 1 1 1
1 1 1 1\\

13 & 1.000 & 98.39 -104.50 36.76 175.29 99.72 -62.70 -120.28 77.66
46.32 37.25 33.31 13.03 -21.19  & 0.39 1.13 1.56 1.35 0.31 1.38 0.65
1.01 1.40 1.05 0.76 0.45 0.15 \\

20 & 0.988 & 5.46 -20.64 -40.09 -28.29 -36.87 -22.32 43.55 172.33 175.48
93.05 -34.38 -55.14 122.29 -158.74 -45.56 89.39 -79.32 136.14 -46.09
134.19  & 0.80 0.80 0.87 1.20 1.18 0.80 0.80 0.80 1.20 0.94 1.20 1.20
1.20 0.80 0.84 1.20 0.88 0.86 1.20 0.92 \\

33 & 0.982 & 18.60 -100.83 161.00 33.34 -79.52 130.05 30.10 -122.23
-55.99 168.40 98.19 94.49 -77.28 4.82 -167.74 65.06 168.02 -28.00 9.50
90.79 -82.85 -3.32 -94.82 -114.72 -71.90 130.07 -169.00 -162.73
-107.41 -86.53 -48.03 -41.65 -14.85  &  1 1 1 1 1 1 1 1 1 1 1 1 1 1 1
1 1 1 1 1 1 1 1 1 1 1 1 1 1 1 1 1 1  \\

42 & 1.000 & -174.51 158.30 -126.60 139.94 -128.83 -149.15 51.30 -135.17 82.97 -31.20 139.69 -1.60 -148.26 28.75 -19.38 27.63 -21.57 35.47 143.15 -50.60 53.19 133.13 -78.68 -119.40 -72.44 103.84 72.66 40.87 -103.49 89.89 -10.03 -55.58 -170.31 93.54 -141.04 136.35 54.50 -23.15 -148.32 27.18 19.58 -125.25  & 0.30 0.28 0.32 0.42 0.28 0.64 0.47 0.72 0.77 0.45 0.39 1.29 0.53 1.09 1.16 1.18 1.76 1.62 0.79 1.02 1.27 1.90 1.72 1.50 1.34 1.85 1.08 1.73 1.31 0.35 1.07 0.84 0.80 0.73 0.68 0.41 0.53 0.64 0.29 0.20 0.24 0.14 \\

\hline
\end{tabular*}
\end{table*}




\section{Conclusions}

Estimator mean and variance was derived for matched and mismatched
filter target power estimators in the case of an arbitrary target. It
was seen that it is sufficient to minimize thermal noise entering the
filter. It was also noted that matched filter estimator contains bias
and often results in larger estimator variance than the mismatched
filter when the target is distributed. The obtained equations for
estimator variance can be used for more specific radar design problems
where there is prior information of the range and power extent of the
target.

In order to search for optimal mismatched filter estimator codes, a
heuristic constrained random local improvement algorithm was used to
find transmission codes that are in many cases extremely close to
theoretical optimum. The width of the estimator variance is inversely
proportional to $\SNR$ and transmission power, and thus the largest
improvements in comparison to binary phase codes can be found for
short transmission codes and poor $\SNR$ values. For good $\SNR$
levels and longer codes, the improvement is not as dramatic.

\section{Future Work}

In this study, we restricted ourselves to targets that do not have
Doppler, and thus the performance of these codes in the presence of
Doppler is not known. The next logical step would be to study
estimation of targets with Doppler. In these cases the optimal
transmission codes may be different. We only studied the
performance of two natural and commonly used linear target power
estimators. A more superior method would be to study target estimation
as a statistical problem, selecting codes that minimize the posterior
distribution of the target variable, given the measurements and prior
information about the target.

\section{Acknowledgements}

The authors acknowledge support of the Academy of Finland through the
Finnish Centre of Excellence in Inverse Problems Research.

\end{document}